\documentclass[useAMS,usenatbib]{mn2e}
\usepackage{amssymb,amsmath}
\usepackage{graphicx}
\usepackage{natbib}
\usepackage{journal_shortcuts}
\newcommand{\bi}{\begin{itemize}}
\newcommand{\ei}{\end{itemize}}
\newcommand{\be}{\begin{equation}}
\newcommand{\ee}{\end{equation}}

\title[Shock heating by FR I radio sources in galaxy clusters]%
{Shock heating by FR I radio sources in galaxy clusters}

\author[M. Br\"uggen et al.]{M.~Br\"uggen$^{1}$, S.~Heinz$^{2}$, E.~Roediger$^{1}$, M.~Ruszkowski$^{3}$, A. Simionescu$^{4}$ \\
$^1$ Jacobs University Bremen, P.O. Box 750\,561, 28725 Bremen,
Germany\\
$^2$ Astronomy Department, University of Wisconsin-Madison, Madison, WI 53706 USA\\
$^3$ Max-Planck-Institut f\"ur Astrophysik, Karl-Schwarzschild-Strasse 1, 85741
Garching, Germany\\
$^4$ Max-Planck-Institut f\"ur Extraterrestrische Physik, Giessenbachstr, 85748 Garching, Germany
}

\begin{document}

\date{Accepted. Received; in original form }

\pagerange{\pageref{firstpage}--\pageref{lastpage}} \pubyear{2005}

\maketitle

\label{firstpage}

\begin{abstract}
  Feedback by active galactic nuclei (AGN) is frequently invoked to
  explain the cut-off of the galaxy luminosity function at the bright
  end and the absence of cooling flows in galaxy clusters. Meanwhile,
  there are recent observations of shock fronts around radio-loud AGN.
  Using realistic 3D simulations of jets in a galaxy cluster, we
  address the question what fraction of the energy of active galactic
  nuclei is dissipated in shocks. We find that weak shocks that
  encompass the AGN have Mach numbers of 1.1-1.2 and dissipate at
  least 2\% of the mechanical luminosity of the AGN. In a realistic
  cluster medium, even a continuous jet can lead to multiple shock
  structures, which may lead to an overestimate of the AGN duty cycles
  inferred from the spatial distribution of waves.
\end{abstract}


\section{Introduction}

There is increasing evidence from deep X-ray observations for shock
waves from radio-loud active galactic nuclei (AGN) in the central
cores of galaxy clusters (e.g. \cite{mcnamara:05}). Examples are the
(weak) shocks in M87 (\citealt{forman:05}, \citealt{simionescu:07}),
Hydra A (\citealt{nulsen:06}) and Perseus (\citealt{fabian:06}).
Using deeper {\sc Chandra} data of the Virgo cluster, \cite{forman:06}
confirmed the presence of a weak shock at 14 kpc and determined its
Mach number as $M\sim 1.2$. This shock was confirmed spectroscopically
using XMM-Newton data by \cite{simionescu:07}. \cite{sanders:06}
report isothermal shocks in Abell 2199 and 2A~0335+096 with Mach
numbers of $\sim 1.5$. Shock waves have also been detected in the
periphery of the low-power isolated radio-galaxy NGC~3801
(\citealt{croston:07}) and a strong shock has been associated
with the expanding FR~I radio galaxy in Centaurus A
(\citealt{kraft:03}).\\

The detection of shocks around FR~I sources has led to the
suggestion that most, if not all, radio galaxies go through a phase
that is associated with shock heating by a supersonically
expanding radio source. \cite{croston:07} have estimated that the
energy stored in the shocked shell is equivalent to the thermal energy
within $\sim$ 11 kpc of the galaxy centre and a factor 25 larger than
the inferred $p\;dV$ work required to inflate the radio lobes. This
suggests that in the early phases of radio-source evolution, the
energy transfer from the AGN to its environment is dominated by shock
heating. 

\cite{nulsen:06} made the following estimate for the heat input into
the ICM by shock waves: The heat per unit mass generated by a shock is
given by \be \Delta Q \sim T \Delta S = E\Delta \ln p/\rho^\gamma ,
\ee

where $E$ is the specific thermal energy, $p$ the pressure and $\rho$
the density of the gas. Thus, the fraction of the thermal energy that
is dissipated, $\Delta Q/E$, is given by the jump of $\ln
p/\rho^\gamma$ in the shock. Three weak shocks are visible in the
X-ray image of M87. For the innermost shock at $\sim$ 3.7 kpc a Mach
number of 1.4 has been inferred, which implies a heat input of $\Delta
Q/E \approx 0.022$ and a shock age of $2.4 \times 10^6$
yrs. Obviously, the heat input of this shock is tiny. However, two
more shocks have been identified at larger radii that require several
times more energy. Thus, a shock of comparable strength to the 3.7 kpc
shock may well occur every $\sim 2.5 \times 10^6$ y. The cooling time
of the gas at 3.7 kpc is $\approx 2.5 \times 10^8$ yrs, so that there
is time for $\sim 100$ such shocks during the cooling time. Therefore,
the combined heat input from $\sim 100$ of these shocks is more than
enough to make up for radiative losses from the gas.\\

In this Letter, we investigate what fraction of the jet energy is
dissipated in shocks around the supersonically expanding radio
source. Using a hydrodynamical simulation of jets in a realistic
cluster set-up and a shock finding algorithm, we quantify the
properties of the shock and the effect on the intracluster medium.

\section{Method}

The initial conditions of our simulation are based on a rerun of the
S2 cluster from \cite{springel:01}, whose properties are sufficiently
close to a typical, massive, X-ray bright cluster with a mass of
$M\sim 7\times 10^{14} M_{\odot}$ and a central temperature of 6
keV. The cluster appears as a classical, relaxed cooling flow cluster
in X-rays. Its density rises steeply in the centre, and the profile is
very similar to the density profiles reported in
\cite{vikhlinin:06}. The setup is the same as used in
\cite{heinz:06}. The output of the {\tt GADGET} SPH simulation serves
as the initial conditions for our simulation.  We use the {\tt FLASH}
code \citep{fryxell:00} which is a modular block-structured adaptive
mesh refinement code, parallelised using the Message Passing
Interface. It solves the Riemann problem on a Cartesian grid using the
Piecewise-Parabolic Method. Our simulation includes $7\times 10^5$
dark matter particles. For the relatively short physical time of the
jet simulation (25 Myrs), radiative cooling and star formation are
neglected, though they were included in the constitutive SPH
simulation.

The computational domain is a $2.8$ Mpc$^3$ box around the cluster's
centre of mass. The maximum resolution at the grid centre corresponds
to a cell size of $174$ pc, implying 11 levels of refinement. The
simulations presented in this letter were performed assuming an
adiabatic equation of state with a uniform adiabatic index of
$\gamma=5/3$.\\

The jet is injected through a nozzle placed at the centre of the
gravitational potential, coincident with the gas density peak of the
central elliptical galaxy.  The nozzle is modeled as two circular
back-to-back inflow boundaries $2$ kpc or 12 resolution elements in
diameter.  The nozzle faces obey inflow boundary conditions fixed by
the jet's mass-, momentum-, and energy fluxes.  This treatment avoids
the entrainment of cluster gas into the jet which is unavoidable in
simpler schemes where the jet is approximated by injecting mass,
momentum, and energy into a finite volume of the cluster that contains
thermal gas and is part of the active computational grid.  We were
thus able to separate cleanly jet fluid and cluster fluid in order to
study the heat input into the ICM only. The jet is centred on the
gravitational potential of the cluster and follows the (slow) motion
of the cluster through the computational domain.

The jet material is injected equally in opposite directions with
velocity $v_{\rm jet}=3\times 10^{9}\,{\rm cm\,s^{-1}}$ and an
internal Mach number of 32.  The jet power of the simulation presented
in this letter was chosen to be $W_{\rm jet}=3\times 10^{45}\,{\rm
ergs\,s^{-1}}$, corresponding to a rather powerful source. Comparing
this luminosity with the sample of cavity systems studied by
\cite{birzan:04}, the jet power in our simulation is at the extreme
end of those observed and well above that cited for M87
(\cite{allen:06}). However, even such powerful sources assume FR~I
morphologies in dense clusters such as this one (see, e.g., Perseus
A). Even though Perseus A is inferred to have a mean power output of
$\sim 10^{44}$ erg s$^{-1}$ \citep{sanders:07}, its peak luminosities
are likely to be significantly larger. Hercules A has a power of
$1.6\times 10^{46}$ erg s$^{-1}$
as implied by the large-scale shocks found around it \citep{nulsen:05,
mcnamara:05} but still exhibits an FR~I/II morphology. In our
simulation, we chose such a high luminosity to ensure that the
jet is able to push through the dense gas of the central galaxy.

\begin{figure*}
\includegraphics[trim=0 80 0 0,clip,width=0.45\textwidth]{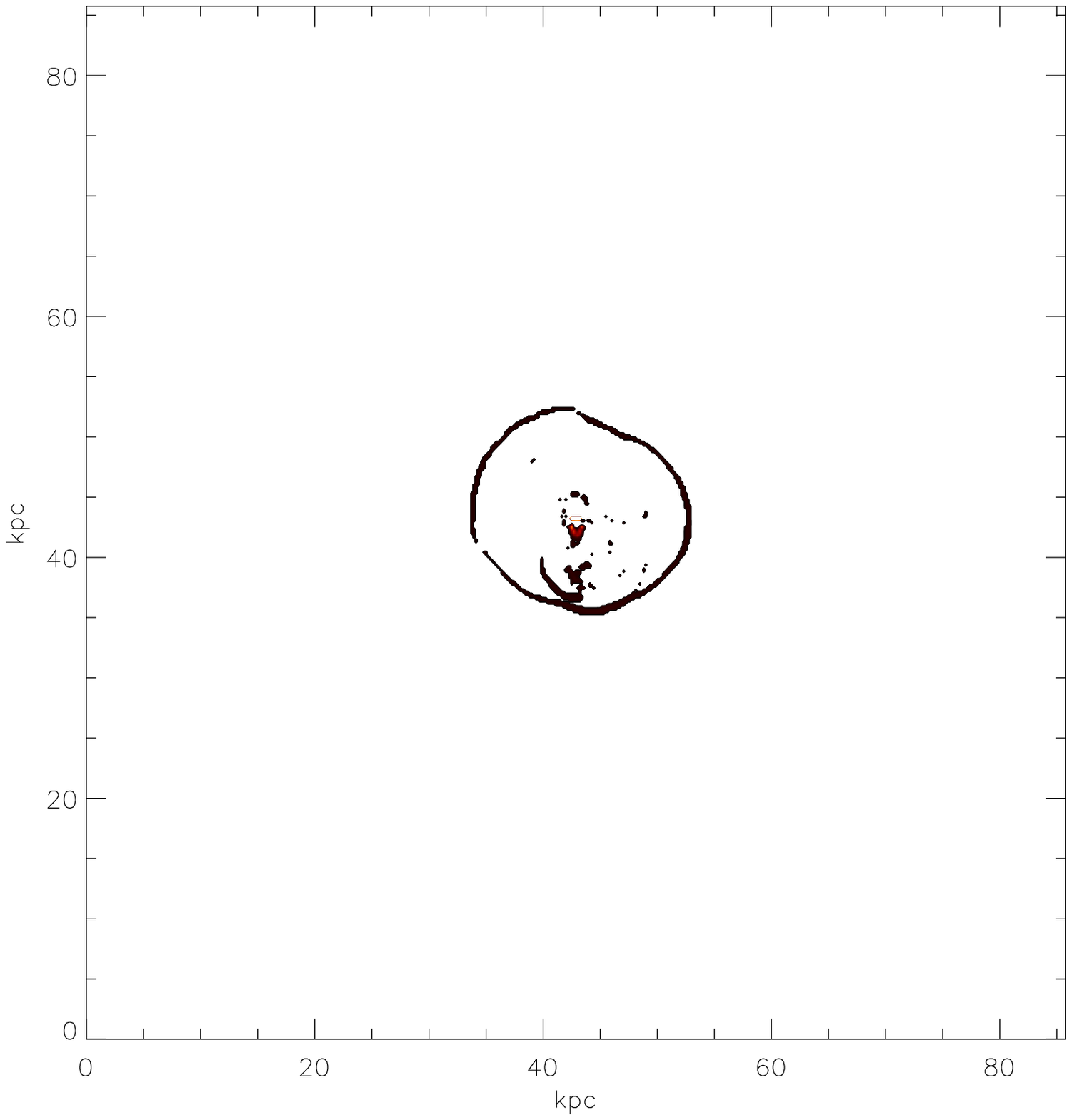}
\includegraphics[trim=0 80 0 0,clip,width=0.45\textwidth]{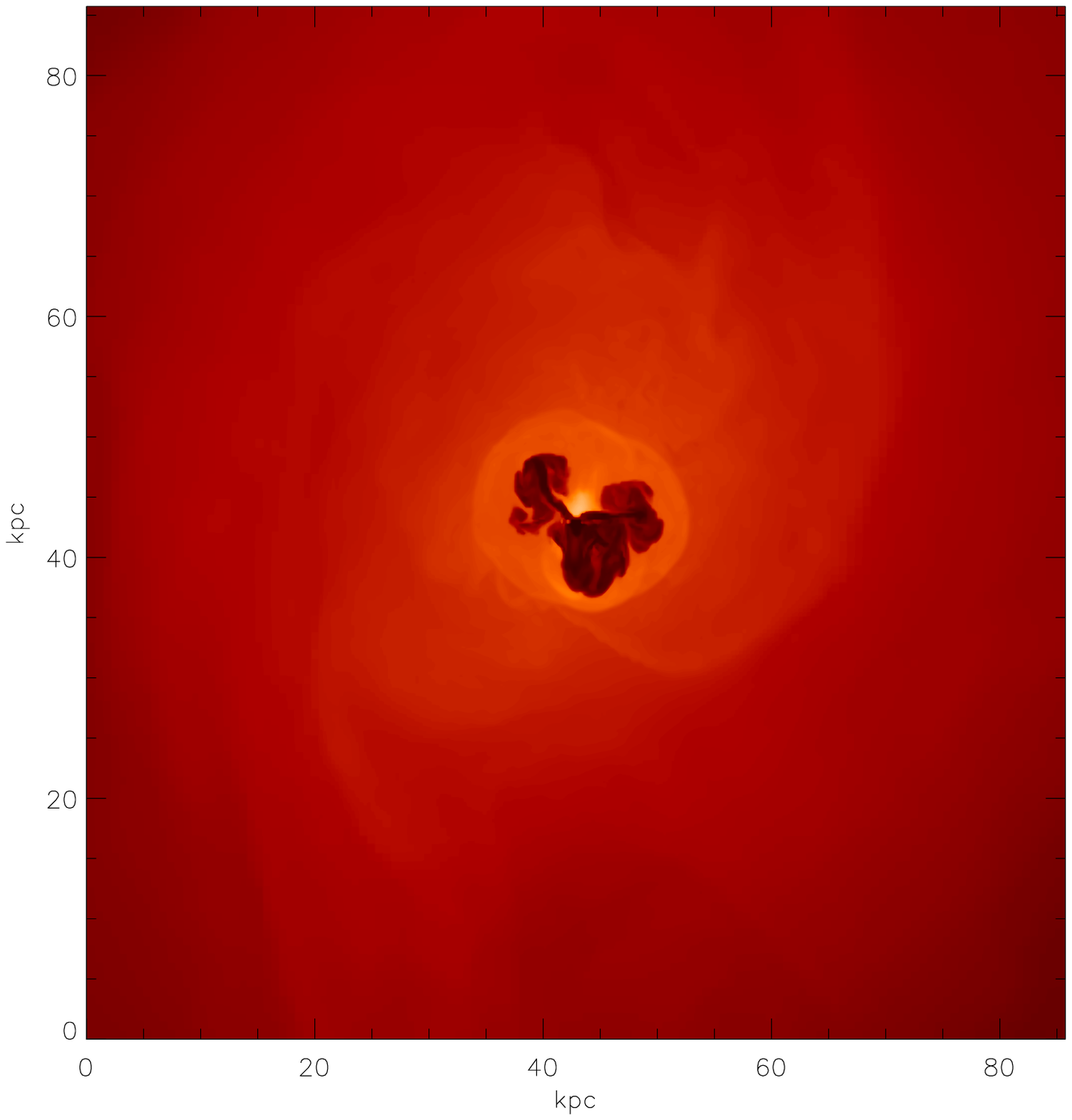}\newline
\includegraphics[trim=0 80 0 0,clip,width=0.45\textwidth]{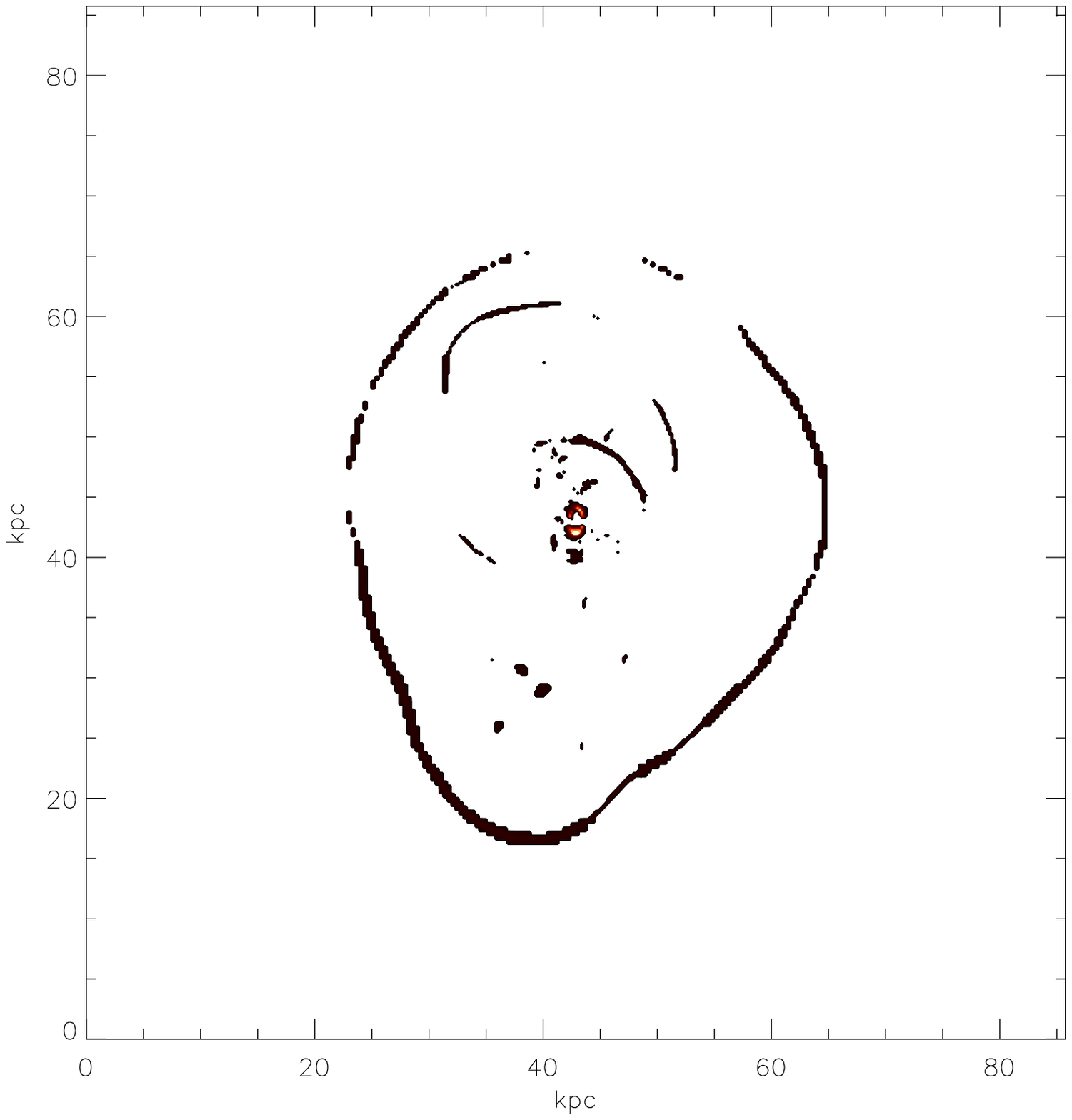}
\includegraphics[trim=0 80 0 0,clip,width=0.45\textwidth]{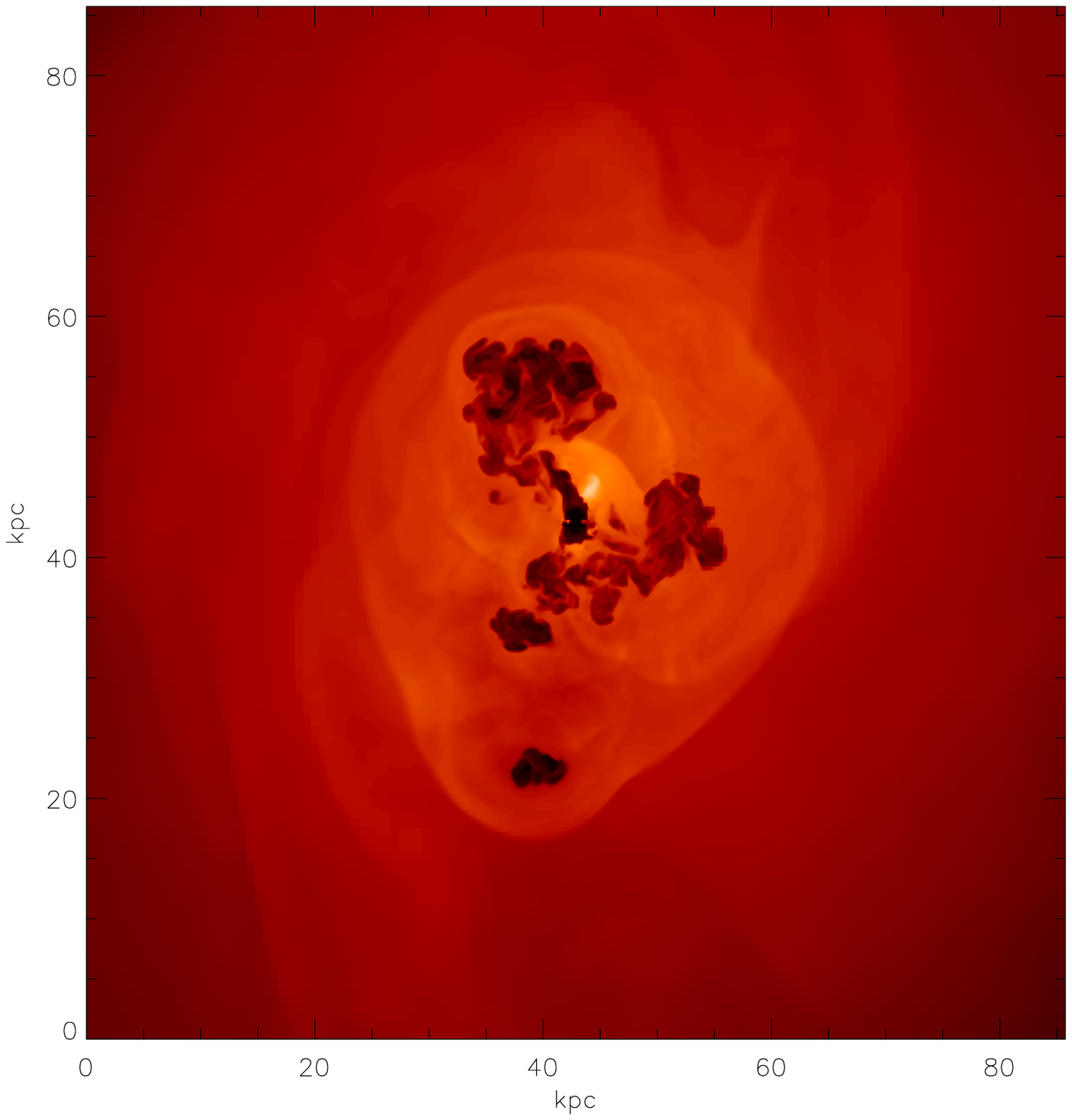}

\caption{Slices through the cluster centre showing the shocks (left) and the gas density (right) at 5 Myrs and 15 Myr after the start of the jet. Shown is only the central part of the computational domain. The entire computational domain represents a volume of $2.8$ Mpc$^3$.}
\label{fig:machdensity}
\end{figure*}

The shocks in our simulation are detected using a multidimensional
shock detection module adopted from the sPPM code
\footnote{S.E.Anderson and P.R.Woodward, World Wide Web
http://www.lcse.umn.edu/research/sppm, Laboratory for Computational
Science and Engineering, University of Minnesota (1995).} based on
pressure jumps across the shock. The basic algorithm evaluates the
jump in pressure in the direction of compression (determined by
looking at the velocity field). If the total velocity divergence is
negative and the relative pressure jump across the compression front
is larger than some chosen value ($\Delta p/p \ge 0.25$), then a zone
is marked as shocked. Using the jumps across the shock in the 3
velocity components, we get the $x$-, $y$-, and $z$-components of a
unit vector pointing in the direction of the velocity jump, hence in
the direction normal to the shock front.  We now project the pre- and
post-shock velocities onto the shock normal. The upstream and
downstream pressure, the upstream velocity and density are then
written out. We have tested this shock detection algorithm with one-
and two-dimensional shock tube problems and found that the jumps in
pressure and density are reproduced very well. The shock structures in
a slice through the central regions of our computational domain are
shown in Fig.~\ref{fig:machdensity}.

\section{Results}

We have simulated a jet that resembles a FR I source in a realistic
cluster environment. Our simulation reproduces the shock structure in
the inner $\sim 100$ kpc of the cluster around an AGN with FR~I
morphology. One can see how the jet inflates bubbles that break off
and start to rise through the cluster medium. This morphology
resembles many of the low-power AGN that are observed to inflate
bubbles at the centres of cooling flow clusters.\\

Fig.~\ref{fig:machdensity} reveals two kinds of shocks: At the working
surface of the jet, i.e. at the point where the jet impacts the
ambient medium, there is a very strong shock wave. In the first
million years after the start of the jet, this shock wave has Mach
numbers with respect to the ICM of $> 30$. The normal of this shock
surface is equal or close to the direction of the jet and does not
encompass the entire jet region, contrary to the outer shock that is
described below. Later on, as the ICM near the jet gets hotter, the
Mach number of the strong inner shock decreases to close to $\sim
8-10$. When motions of the ambient medium cause the jet to break off
or to change direction, this shock can detach and a new working
surface forms. As the jet jitters and ambient material moves into the
jet, multiple shock fronts develop in the region close to the
AGN. This leads to multiple shock features, as one can see in the
botton left panel of Fig.~\ref{fig:machdensity}. In our simulation, we
see that a {\it continuous} jet can lead to multiple shock fronts as those
observed in M87 and other clusters. Hence, the exitence of multiple
shock fronts does not necessarily imply an intermittency of the AGN.\\

The second kind of shock is a weak and nearly spherical shock that
travels from the point of the injection region outwards through the
cluster.  The Mach number of the outer shock remains at fairly
constant values of around 1.1 - 1.2 for the largest part of its
propagation through the core of the cluster. The pressure jump across
a shock of $M=1.2$ is 1.55. The outer shock is a pressure wave that is
driven by the additional pressure from the injected gas in the core of
the cluster. This is different from the strong, inner shock that is
driven by the ram pressure of the jet. After about
10 Myrs the outer shock becomes prolate in the direction of the
jet. As the bubbles rise mainly in the jet direction, the pressure
also increases preferentially in the direction of the jet. The Mach
number is slightly higher in the direction of the jet than
at the sides of the outer shock front.\\

\begin{figure}
\centering\resizebox{\hsize}{!}%
{\includegraphics[angle=0]{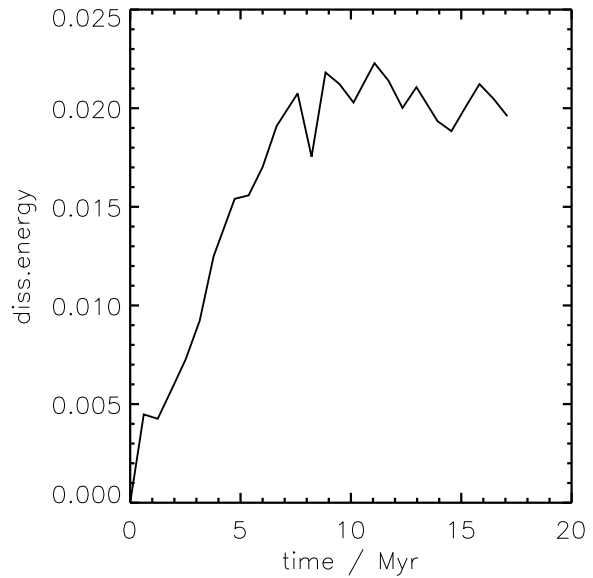}}
\caption{Plot of the energy generated at the outer shock surface per time divided by the mechanical luminosity of the jet.}
\label{fig:shockenergy}
\end{figure}

Next, we wish to compute the total energy thermalised in the shock
front. One can write for the thermal energy flux generated at the shock:

\begin{equation}
F=\left[e_{\rm d}-e_{\rm u}(\rho_{\rm d}/\rho_{\rm u})^{\gamma}\right]v_{\rm d} ,
\end{equation}
where the subscripts d and u denote down- and upstream quantities,
respectively, $e$ is the thermal energy density, $\rho$ gas density
and $v$ velocity. The second term inside the brackets subtracts the
effect of adiabatic compression suffered at a shock. The total
thermalised energy input per time (by shocks) divided by the
mechanical luminosity of the jet is shown in
Fig.~\ref{fig:shockenergy}. About 2\% of the mechanical luminosity of
the jet are converted lastingly into internal energy. The properties
of the outer shock are found to be relatively insensitive to the
mechanical luminosity of the jet. The inner, strong shocks are much
more efficient at generating energy because their Mach number is much
higher. However, their area is relatively small and thus they may be
important for the interstellar medium of the host galaxy, but they are
unlikely to have a significant effect on the thermal state of the
ICM. The properties of the inner shock depend also quite sensitively
on the exact jet parameters and to some degree on the numerical
resolution of our grid. Meanwhile, the properties of the outer shock
are not sensitive to the numerical resolution and appear converged. We
note that the timescale and geometry of the initial energy release is
important for the computation of the energy deposited in the ICM or
interstellar medium. If the same total energy were injected in the
form of thermal energy in pressure equilibrium with the surroundings,
the amount of energy transferred to the ICM in the form of shocks
would be very different. Even taking into account the uncertainties of
our jet model, the approach presented here is much more realistic than
schemes in which the energy is injected in pressure equilibrium.

\begin{figure}
\centering\resizebox{\hsize}{!}%
{\includegraphics[angle=0]{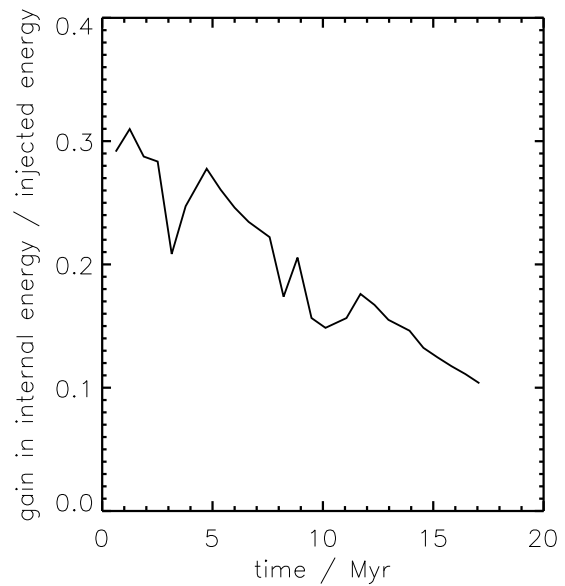}}
\caption{Increase in internal energy of the ambient medium
(i.e. excluding the jet material) within the outer shocks divided by
the total mechanical energy injected so far.}
\label{fig:dissenergy}
\end{figure}

The total increase in internal energy per unit time of the ambient
medium (i.e. excluding the jet material) within the outer shock
divided by the total mechanical power of the jet is shown in
Fig.~\ref{fig:dissenergy}. We see that for the first 15 Myrs of the
AGN activity, a bit more than 30~\% of the injected energy has been
converted to thermal energy in the inner core. This is much more than
what has been thermalised by the outer shock. The temperature increase
in the cluster core is mainly caused by $pdV$ work from the expanding
bubbles. While the shocks thermalise only a few per cent of the jet
energy, this also raises the entropy of the cluster.  On the other
hand, $pdV$ work by the expanding bubbles is adiabatic until the gas
motions induced by the rising bubbles are dissipated by viscous
processes (\citealt{nulsen:06}). Yet, a succession of shocks can be
sufficient to offset the radiative cooling of the entire ICM, as only
4-5 shock fronts that are permanently present lead to a conversion of
$\sim$ 10 \% of the jet energy.\\

Clearly, magnetic fields can alter the dynamics of the jet and the
radio lobes. As shown in \cite{ruszkowski:07}, large-scale cluster
magnetic fields tend to drape themselves around the rising bubbles and
can suppress the fragmentation of the bubbles. The ensuing dynamics
can be different from the dynamics modelled here, but the outer shock
structure is unlikely to be substantially different. In the presence
of a physical viscosity with values close to the Spitzer value, the
dissipation of the outer, weak shocks can be higher than what we find
in our simulation (see also \citealt{bruggen:05,ruszkowski:04b}).

We have presented a single simulation of a shock that has been
produced by a FR~I jet. Obviously, there are a lot of parameters
that can be varied. However, these kind of simulations are very expensive; on
NAS Columbia a single simulation took more than 12,000 CPU hours (on
64 processors). Hence, we chose one exemplary case that reproduces
properties that match observations such as those in Hercules or
M87 (despite wide discrepancies in jet powers). In the long
term, a proper parameter study ought to be conducted.

\section*{Acknowledgements}

The anonymous referee is thanked for a helpful report.
MB and ER acknowledge the support by the DFG grant BR 2026/3 within
the Priority Programme ``Witnesses of Cosmic History'' and the
supercomputing grants NIC 2195 and 2256 at the John-Neumann Institut
at the Forschungszentrum J\"ulich. SH acknowledges support
through NASA grant TM5-6007X. We also acknowledge a
supercomputing grant on NAS Columbia.

The results presented were produced using the FLASH code, a product of the DOE
ASC/Alliances-funded Center for Astrophysical Thermonuclear Flashes at the
University of Chicago.

%
\bibliographystyle{mn2e}
\bibliography{%
../BIBLIOGRAPHY/radio,%
../BIBLIOGRAPHY/metals,%
../BIBLIOGRAPHY/shbib,%
../BIBLIOGRAPHY/marcus%
}

\bsp

\label{lastpage}

\end{document}